\documentclass{amsart}

\usepackage{epsfig}
\usepackage{graphicx}
\usepackage{amsmath,amssymb,amsfonts,latexsym}
\usepackage{graphicx}

\usepackage{amssymb}
\usepackage{enumerate, xspace}
\usepackage{dsfont}
\usepackage{afterpage}
\usepackage{changebar}
\usepackage{amsthm}
\usepackage{rotating}

\numberwithin{equation}{section}

\begin{document}

\title[Einstein-Podolski-Rosen paradox]
{Einstein-Podolski-Rosen paradox, Non-commuting operator, Complete wavefunction and Entanglement}
\author {Andrew Das Arulsamy}
\address{Condensed Matter Group, Division of Interdisciplinary Science, F-02-08 Ketumbar Hill, Jalan Ketumbar, 56100 Kuala-Lumpur, Malaysia}
\email{sadwerdna@gmail.com}


\date{\today}

\begin{abstract}
Einstein, Podolski and Rosen (EPR) have shown that any wavefunction (subject to the Schr$\ddot{\rm o}$dinger equation) \textit{can} describe the physical reality completely, and any two observables associated to two non-commuting operators \textit{can} have simultaneous reality. In contrast, quantum theory claims that the wavefunction \textit{can} capture the physical reality completely, and the physical quantities associated to two non-commuting operators \textit{cannot} have simultaneous reality. The above contradiction is known as the EPR paradox. Here, we unambiguously expose that there is a hidden assumption made by EPR, which gives rise to this famous paradox. Putting the assumption right this time leads us not to the paradox, but only reinforces the correctness of the quantum theory. However, it is shown here that the entanglement phenomenon between two physically separated particles (they were entangled prior to separation) can only be proven to exist with a `proper' measurement. 
\end{abstract}

\maketitle

\section{Introduction}

In a classical sense, a theory's sole purpose is to help us picture the physical reality correctly and completely such that there is a crystal clear distinction between the theory and the physical reality~\cite{epr}. The distinction here means that there is no interaction between any observable and the experimenter. Based on this viewpoint, Einstein, Podolski and Rosen (EPR) moved on to show that the quantum theory is incomplete because there is no one-to-one correspondence between the physical theory and the reality. Moreover, the EPR paradox actually render the quantum theory to be downright incorrect~\cite{david}---because any two non-commuting operators (that may exist in the Schr$\ddot{\rm o}$dinger equation) can lead to two distinct observables having simultaneous reality. In other words, EPR claims that if the wavefunction is to be considered complete, then this completeness contradicts with the notion that two physical quantities (belonging to two non-commuting operators) not having simultaneous reality. The reason for the above contradiction is because the quantum theory states that the wavefunction is complete, and the physical quantities associated to two non-commuting operators indeed cannot have simultaneous reality.  

Here, we will re-evaluate this paradox and tackle it head-on, without invoking any additional indirect notion implied from this paradox, namely, the observer-observable interaction and the entanglement (interaction between two observables) phenomenon, which were first proposed by Bohr~\cite{bohr} and Bohm~\cite{bohm}, respectively. Our strategy here is to go back in time and re-evaluate the paradox in its original framework in order to expose the existence of an \textit{ad hoc} assumption. This assumption is shown to be logically incorrect, and after putting it in the proper context, we will find that the EPR arguments to be in complete agreement with the quantum theory. 

In quantum theory, one can assign a unique operator to each physical quantity, namely, the total energy corresponds to the Hamilton operator, $\textbf{H} = -(\hbar^2/2m)\Delta + \textbf{V}$ where $-(\hbar^2/2m)\Delta = \textbf{p}^2/2m$ and $\textbf{V}$ are the kinetic and potential (including other interaction) energy operators, respectively, where $m$ is the mass of a particle, $\hbar = h/2\pi$, $h$ denotes the Planck constant, and $\Delta = \nabla^2$. Moreover, $\textbf{p} = -i\hbar\nabla$ and $\textbf{r}$ are the respective momentum and position operators. Here, we will deal only with $\textbf{p}$ and $\textbf{r}$ because they are sufficient. An operator can operate on an eigenfunction, $\psi$ such that
\begin {eqnarray}
\textbf{p}\psi = p\psi, \label{eq:1}
\end {eqnarray}  
\begin {eqnarray}
\textbf{r}\psi = r\psi, \label{eq:2}
\end {eqnarray}  
where $p$ and $r$ are the momentum and position eigenvalues for the particle in the eigenstate $|\psi\rangle$. These eigenfunctions can also be referred as wavefunctions. Since $\textbf{r}$ and $\textbf{p}$ do not commute,  
\begin {eqnarray}
[\textbf{r},\textbf{p}] = \textbf{r}\textbf{p} - \textbf{p}\textbf{r} = i\hbar = \sqrt{-1}\hbar, \label{eq:3}
\end {eqnarray}  
we are then forced (not in a bad way) to adopt the Copenhagen interpretation, and conclude both the momentum and position cannot be observed simultaneously because measuring any one of them ($\textbf{p}$ or $\textbf{r}$), will alter the eigenfunction in such a way that it is never possible to know the other ($\textbf{r}$ or $\textbf{p}$) simultaneously. \textit{Nota bene}, the commutation relation given in Eq.~(\ref{eq:3}) is actually serious due to $\sqrt{-1}$.   

Let us first recall the original EPR criterion for completeness, which reads---\textit{if, without in any way disturbing a system, we can predict with certainty (i.e. with probability equal to unity) the value of a physical quantity, then there exists an element of physical reality corresponding to this physical quantity} or \textit{every element of the physical reality must have a counterpart in the physical theory}~\cite{epr}. The above-stated EPR criterion is used by EPR to reason that both $p$ and $r$ are real and definite from Eqs.~(\ref{eq:1}) and~(\ref{eq:2}) representing the momentum and the position of a particle in the eigenstate $|\psi\rangle$. This means that, EPR implicitly assumed $\psi$ to represent any function such that $\textbf{p}\psi^{(\alpha)} = p\psi^{(\alpha)}$, $\textbf{r}\psi^{(\beta)} = r\psi^{(\beta)}$ and $\psi^{(\alpha)} \neq \psi^{(\beta)}$. In this work, we will show that this hidden assumption ($\psi^{(\alpha)} \neq \psi^{(\beta)}$) is false by considering an atomic hydrogen.    

Apart from that, in view of the above simultaneous reality (both $p$ and $r$ are simultaneously observable), EPR gave us two options~\cite{epr}---either (1) the description derived from the quantum theory based on the wavefunction ($\Psi$) is incomplete, or (2) if any two physical quantities do not have simultaneous reality, then these quantities correspond to two non-commuting operators. The Copenhagen interpretation endorses (2) such that the wavefunction is to be considered complete. On the contrary, EPR actually `showed' that the negation of (1) indeed leads to the negation of (2) and thus they concluded that the quantum theory has got be incomplete (if not downright incorrect) because it is clearly not self-consistent. In fact, EPR proved that when statement (1) is false, then (2) is also false, which means, the negation of (1) leads to the negation of (2). This \textit{is} the contradiction, which is known as the EPR paradox. For example, the original EPR option reads---\textit{when the operators corresponding to two physical quantities do not commute the two quantities cannot have simultaneous reality}~\cite{epr}, which can also be written as, \textit{when two physical quantities do not have simultaneous reality, then their operators corresponding to these two physical quantities do not commute}, which is nothing but (2) given above. Here, we will show that the negation of (1) does not lead us to the negation of (2).  

Of course, experimentally we did not find any serious violation against the quantum theory~\cite{aspect,gw}. But we, or at least some of us (in the post-Copenhagen-interpretation era) `feel' the negation of (1) should not negate (2) in the first place, or at least logically the quantum theory should be self-consistent with itself, without the need to invoke the experiments as the final judge. Here, we will re-evaluate the double negation [(1) and (2)] and we will find that the EPR paradox ceases to exist. Prior to conclusions, we will discuss the notion of entanglement due to Bohm~\cite{bohm} because it is the only surviving issue originating from the EPR paradox. In summary, our aims here are to show that (i) the EPR hidden assumption ($\psi^{(\alpha)} \neq \psi^{(\beta)}$) is false and (ii) the complete wavefunction does not lead to simultaneous reality for both $p$ and $r$.     
          
\section{Incomplete and complete wavefunctions}

We reconsider the exact systems (I and II) studied by EPR. We let the two systems to interact between time, $t = 0$ to $t = T$ such that there were no interaction between system I and II when $t < 0$ (before interaction) and $t > T$ (after interaction). However, the Schr$\ddot{\rm o}$dinger equation,
\begin {eqnarray}
i\hbar\frac{\partial \Psi}{\partial t} = \textbf{H}\Psi, \label{eq:sc1}
\end {eqnarray}  
cannot be used to determine the states of the combined system (I + II) after the interaction ($t > T$), even if we knew the states of the two systems before they interact or when $t < 0$. This unfortunate scenario is due to the quantum theory, and the reason is explained below. 

After the interaction, following EPR, we can let $a_1, a_2, \cdots$ as the eigenvalues corresponding to an operator $\textbf{A}$, which is related to a physical quantity $A$. In system I, we can also denote the eigenfunctions (related to $A$) as $u_1(x_{\rm I}), u_2(x_{\rm I}), \cdots$ where $x_{\rm I}$ symbolically denotes the collection of variables belonging to system I. The wavefunction for the combined system (I + II), 
\begin {eqnarray}
\Psi(x_{\rm I},x_{\rm II}) = \sum_{n = 1}^{\infty}\psi_n(x_{\rm II})u_n(x_{\rm I}), \label{eq:4}
\end {eqnarray}  
where $x_{\rm II}$ is now a symbolic notation for the variables belonging to system II, and $\psi_n(x_{\rm II})$ are the coefficients associated to the orthogonal eigenfunctions, $u_n(x_{\rm I})$ such that one obtains an expanded series for $\Psi$. We now measure $A$, and obtained $a_k$. This means that, after the measurement, the first system is in the eigenstate $|u_k(x_{\rm I})\rangle$, while system II is in the eigenstate $|\psi_k(x_{\rm II})\rangle$. As a consequence, the wavefunction given in Eq.~(\ref{eq:4}) is said to have `collapsed' into   
\begin {eqnarray}
\Psi_k(x_{\rm I},x_{\rm II}) = \psi_k(x_{\rm II})u_k(x_{\rm I}). \label{eq:5}
\end {eqnarray}  
Recall that the set of orthogonal eigenfunctions, $u_n(x_{\rm I})$ only refer to a physical quantity $A$. Subsequently, we can now measure (also in system I) another physical quantity, $B$ associated to an operator $\textbf{B}$ with $b_1, b_2, \cdots$ as eigenvalues, and these eigenvalues correspond to a new set of orthogonal eigenfunctions, $v_1(x_{\rm I}), v_2(x_{\rm I}), \cdots$. In this case, Eq.~(\ref{eq:4}) reads
\begin {eqnarray}
\Psi'(x_{\rm I},x_{\rm II}) = \sum_{s = 1}^{\infty}\varphi_s(x_{\rm II})v_s(x_{\rm I}), \label{eq:6}
\end {eqnarray}  
where $\varphi_s(x_{\rm II})$ is now the coefficients associated to $v_s(x_{\rm I})$. After measuring $B$, we obtained $b_r$, and consequently, the wavefunction given in Eq.~(\ref{eq:6}) collapsed into     
\begin {eqnarray}
\Psi_r(x_{\rm I},x_{\rm II}) = \varphi_r(x_{\rm II})v_r(x_{\rm I}). \label{eq:7}
\end {eqnarray}  
Equation~(\ref{eq:7}) means that system I is now in the eigenstate $|v_r(x_{\rm I})\rangle$, while system II is in the eigenstate $|\varphi_r(x_{\rm II})\rangle$. Obviously, system I can be in two different eigenstates, namely, $|u_k(x_{\rm I})\rangle$ (corresponding to the physical quantity $A$) and $|v_r(x_{\rm I})\rangle$ (corresponding to the physical quantity $B$). Similarly, system II can also be in two different eigenstates, $|\psi_k(x_{\rm II})\rangle$ (due to $A$) and $|\varphi_r(x_{\rm II})\rangle$ (due to $B$). As a result, EPR concluded---one can assign two different wavefunctions, for example, $\psi_k(x_{\rm II})$ and $\varphi_r(x_{\rm II})$ to the same reality (system II). 

\textit{Nota bene}, these two wavefunctions ($\psi_k(x_{\rm II})$ and $\varphi_r(x_{\rm II})$) have been assigned to the same reality (system II) because we had assigned two different wavefunctions ($u_k(x_{\rm I})$ and $v_r(x_{\rm I})$) for the same reality, system I in the first place. In particular, EPR assigned two different wavefunctions ($u_k(x_{\rm I})$ and $v_r(x_{\rm I})$) for system I for two different physical quantities, $A$ and $B$. Consequently, they had to assign two different wavefunctions ($\psi_k(x_{\rm II})$ and $\varphi_r(x_{\rm II})$) for two different physical quantities, $A$ and $B$ in system II. Such assignments mean two things (i) the wavefunctions defined by EPR are incomplete and (ii) there is this notion known as entanglement has been invoked, which was pointed out by Bohm~\cite{bohm}. First we will address the question why the EPR wavefunctions are incomplete by definition, after which we will tackle the entanglement issue. 

If one assigns two different wavefunctions ($u_k(x_{\rm I})$ and $v_k(x_{\rm I})$) to two different physical quantities ($A$ and $B$) for the same reality (system I), then the wavefunctions ($u_k(x_{\rm I})$ and $v_k(x_{\rm I})$) are incomplete by definition. For system II, $\psi_k(x_{\rm II})$ and $\varphi_r(x_{\rm II})$ are also incomplete by definition. The definition for a complete wavefunction is that the wavefunction should remain the same for all measurable physical quantities in a given system. This completeness criterion for the wavefunction is based on the atomic hydrogen. In fact, we do have a complete wavefunction (representing a bounded electron) for atomic hydrogen (a system). We can define this hydrogenic wavefunction as complete because you can solve it analytically, and it remains the same for all physical quantities. We will revisit this atomic hydrogen after studying the EPR analysis on non-commuting operators, $\textbf{AB} - \textbf{BA} \neq 0$.

In view of this completeness criterion for the wavefunctions, we can now suppose $u^{\rm I}_k(x_{\rm I},x_{\rm II})$ and $v^{\rm II}_r(x_{\rm I},x_{\rm II})$ are complete wavefunctions (after the interaction) for system I and system II, respectively, and consequently,   
\begin {eqnarray}
\textbf{A}u^{\rm I}_k(x_{\rm I},x_{\rm II}) = a'_ku^{\rm I}_k(x_{\rm I},x_{\rm II}), ~~~\textbf{B}u^{\rm I}_k(x_{\rm I},x_{\rm II}) = b'_ku^{\rm I}_k(x_{\rm I},x_{\rm II}), \label{eq:8}
\end {eqnarray}  
for system I and
\begin {eqnarray}
\textbf{A}v^{\rm II}_r(x_{\rm I},x_{\rm II}) = a'_rv^{\rm II}_r(x_{\rm I},x_{\rm II}), ~~~\textbf{B}v^{\rm II}_r(x_{\rm I},x_{\rm II}) = b'_rv^{\rm II}_r(x_{\rm I},x_{\rm II}), \label{eq:9}
\end {eqnarray}  
for system II. If we suppose $\psi^{\rm I + II}_w(x_{\rm I},x_{\rm II})$ is the complete wavefunction for the combined system (I + II) during interaction ($0 < t < T$), then 
\begin {eqnarray}
&&\textbf{A}\psi^{\rm I + II}_w(x_{\rm I},x_{\rm II}) = a_w\psi^{\rm I + II}_w(x_{\rm I},x_{\rm II}), ~~~ \textbf{B}\psi^{\rm I + II}_w(x_{\rm I},x_{\rm II}) = b_w\psi^{\rm I + II}_w(x_{\rm I},x_{\rm II}). \label{eq:10}
\end {eqnarray}  
Before the interaction ($t < 0$), Eqs.~(\ref{eq:8}) and~(\ref{eq:9}) read  
\begin {eqnarray}
\textbf{A}u^{\rm I}_y(x_{\rm I}) = a_yu^{\rm I}_y(x_{\rm I}), ~~~\textbf{B}u^{\rm I}_y(x_{\rm I}) = b_yu^{\rm I}_y(x_{\rm I}), \label{eq:11}
\end {eqnarray}  
\begin {eqnarray}
\textbf{A}v^{\rm II}_z(x_{\rm II}) = a_zv^{\rm II}_z(x_{\rm II}), ~~~\textbf{B}v^{\rm II}_z(x_{\rm II}) = b_zv^{\rm II}_z(x_{\rm II}), \label{eq:12}
\end {eqnarray}  
for system I and system II, respectively. Observe that $u^{\rm I}_y(x_{\rm I})$ and $v^{\rm II}_z(x_{\rm II})$ in Eqs.~(\ref{eq:11}) and~(\ref{eq:12}) depend only on $x_{\rm I}$ and $x_{\rm II}$, respectively, whereas, $u^{\rm I}_k(x_{\rm I},x_{\rm II})$ and $v^{\rm II}_r(x_{\rm I},x_{\rm II})$ in Eqs.~(\ref{eq:8}) and~(\ref{eq:9}) need to incorporate the effect of the interaction such that $u^{\rm I}_k(x_{\rm I},x_{\rm II})$ and $v^{\rm II}_r(x_{\rm I},x_{\rm II})$ also require additional variables denoted by $x_{\rm II}$ and $x_{\rm I}$, respectively. 

Apparently, even if we knew the complete wavefunctions ($u^{\rm I}_y(x_{\rm I})$ and $v^{\rm II}_z(x_{\rm II})$) for each particle (I and II, respectively) prior to their interaction, we are unable to predict (\textit{via} the Schr$\ddot{\rm o}$dinger equation) the complete wavefunctions ($u^{\rm I}_k(x_{\rm I},x_{\rm II})$ and $v^{\rm II}_r(x_{\rm I},x_{\rm II})$) for particle I and II, respectively, after the interaction. The reason is that after the interaction, the new complete wavefunctions picked up some additional variables, namely, $x_{\rm II}$ in $u^{\rm I}_k(x_{\rm I},x_{\rm II})$ and $x_{\rm I}$ in $v^{\rm II}_k(x_{\rm I},x_{\rm II})$. Picking up additional variables here means that there is a wavefunction transformation such that the new or transformed wavefunction will look qualitatively and/or quantitatively different from the previous one~\cite{ptp}. Recall here that $x_{\rm I}$ and $x_{\rm II}$ symbolically denote the collection of variables belonging to system I and system II, respectively. Usually, $u^{\rm I}_k(x_{\rm I},x_{\rm II})$ or $v^{\rm II}_k(x_{\rm I},x_{\rm II})$ can also be written as a linear combination, or can be combined nonlinearly as given in Eq.~(\ref{eq:4}) or Eq.~(\ref{eq:6}), respectively. In both cases, it is to be noted here that these or any other combinations should be regarded as educated guesses~\cite{ptp}.   

Regardless whether the wavefunctions are complete or not, if the operators, $\textbf{A}$ and $\textbf{B}$ commute, then these pairs of eigenvalues, $a_k$ and $b_k$ (from Eqs.~(\ref{eq:5}) and~(\ref{eq:7}), respectively), $a'_k$ and $b'_k$ (from Eq.~(\ref{eq:8})), $a'_r$ and $b'_r$ (from Eq.~(\ref{eq:9})), $a_y$ and $b_y$ (from Eq.~(\ref{eq:11})), and $a_z$ and $b_z$ (from Eq.~(\ref{eq:12})) can have simultaneous reality. In the subsequent section, we will show that if the wavefunctions are incomplete, and if the two operators ($\textbf{A}$ and $\textbf{B}$) are non-commuting, then the associated physical quantities ($A$ and $B$) can have simultaneous reality. On the contrary, if the wavefunctions are complete, and the two operators are non-commuting, then the two physical quantities cannot have simultaneous reality, as correctly predicted by the quantum theory. 

\subsubsection{Non-commuting operators}      

Earlier,  we did not bother to restrict the operators, $\textbf{A}$ and $\textbf{B}$ to commute or to not commute. However, following EPR, these two operators can also represent two non-commuting operators corresponding to these eigenfunctions, $\psi_k(x_{\rm II})$ and $\varphi_r(x_{\rm II})$ given above (see Eqs.~(\ref{eq:5}) and~(\ref{eq:7})). If the two systems (I and II) are to be considered as two particles (with continuous spectrum) then one can assign $\textbf{A}$ as the momentum ($\textbf{p}$) operator, and $\textbf{B}$ as the position ($\textbf{r}$) operator. This means that Eq.~(\ref{eq:4}) reads
\begin {eqnarray}
\Psi(x_{\rm I},x_{\rm II}) = \int_{-\infty}^{\infty}\psi_p(x_{\rm II})u_p(x_{\rm I}){\rm d}p = \int_{-\infty}^{\infty}e^{(i/\hbar)(x_{\rm I} - x_{\rm II} + x_{0})p}{\rm d}p, \label{eq:13}
\end {eqnarray}  
where        
\begin {eqnarray}
u_p(x_{\rm I}) = e^{(i/\hbar)x_{\rm I}p},~~~ \psi_p(x_{\rm II}) = e^{-(i/\hbar)(x_{\rm II} - x_{0})p}. \label{eq:14}
\end {eqnarray}  
Here $p$ denotes the momentum (also indicated with a subscript), $x_{\rm I}$ and $x_{\rm II}$ (now) denote the positions of particle I and particle II, respectively. Both eigenfunctions, $u_p(x_{\rm I})$ and $\psi_p(x_{\rm II})$ correspond to the momentum operators $\textbf{p}(x_{\rm I}) = -i\hbar(\partial/\partial x_{\rm I})$ and $\textbf{p}(x_{\rm II}) = -i\hbar(\partial/\partial x_{\rm II})$. Therefore, one obtains $\textbf{p}(x_{\rm II})\psi_p(x_{\rm II}) = -p\psi_p(x_{\rm II})$, which is the momentum for particle II.

For the position operator ($\textbf{r}$), Eq.~(\ref{eq:6}) reads
\begin {eqnarray}
\Psi'(x_{\rm I},x_{\rm II}) = \int_{-\infty}^{\infty}\varphi_x(x_{\rm II})v_x(x_{\rm I}){\rm d}x, \label{eq:15}
\end {eqnarray}  
where        
\begin {eqnarray}
v_x(x_{\rm I}) = \delta(x_{\rm I} - x),~~~ \varphi_x(x_{\rm II}) = \delta(x - x_{\rm II} + x_{0}). \label{eq:16}
\end {eqnarray}
The eigenfunctions, $v_x(x_{\rm I})$ and $\varphi_x(x_{\rm II})$ correspond to position operators, $\textbf{r}(x_{\rm I})$ and $\textbf{r}(x_{\rm II})$, respectively, and therefore, $\textbf{r}(x_{\rm II})\varphi_x(x_{\rm II}) = \textbf{r}(x_{\rm II})\delta(x - x_{\rm II} + x_{0}) = (x + x_{0})\delta(x - x_{\rm II} + x_{0})$. Note here that there are two wavefunctions associated to a single particle (particle II), $\psi_p(x_{\rm II}) = e^{-(i/\hbar)(x_{\rm II} - x_{0})p}$ for the momentum, and $\varphi_x(x_{\rm II}) = \delta(x - x_{\rm II} + x_{0})$ for the position. These wavefunctions are incomplete by definition, which implies that one can indeed obtain simultaneous reality for both $\textbf{p}(x_{\rm II})$ and $\textbf{r}(x_{\rm II})$. In fact, we have warned you earlier on the seriousness of having $i = \sqrt{-1}$ in the commutation relation, which is the reason why we need to construct two different wavefunctions for the non-commuting operators, $\textbf{p}(x_{\rm II})$ and $\textbf{r}(x_{\rm II})$ for the same particle (particle II) such that $i$ is removed, and to obtain two real eigenvalues, $-p$ and $x + x_0$ where both can be observed simultaneously. In summary, we have shown that if the wavefunctions are incomplete, then the physical quantities associated to two non-commuting operators can have simultaneous reality. 

Now, we need to consider a complete wavefunction for a single particle, and check whether this wavefunction can lead the non-commuting momentum and position operators to have simultaneous reality. Although logically, we have shown earlier (due to $i$) that it (simultaneous reality) is never possible. But let us just check it out to be sure. The only wavefunction that is guaranteed to be complete is the wavefunction for atomic hydrogen because this is the only real physical quantum system that can be solved analytically such that the complete wavefunction represents the single bounded electron (depending on its energy level). The hydrogenic wavefunction~\cite{david}   
\begin {eqnarray}
&&\psi_{nlm} = \sqrt{\bigg(\frac{2}{na_{\rm B}}\bigg)^{3}\frac{(n-l-1)!}{2n[(n+l)!]^3}} e^{\frac{-r}{na_{\rm B}}}\bigg(\frac{2r}{na_{\rm B}}\bigg)^{l}\big[L^{2l+1}_{n-l-1}(2r/na_{\rm B})\big]Y^m_l(\theta,\phi), \label{eq:17}
\end {eqnarray}
where $n$, $l$ and $m$ are the usual quantum numbers---principal, azimuthal and magnetic, respectively, $a_{\rm B}$ is the Bohr radius, $L^{2l+1}_{n-l-1}(2r/na_{\rm B})$ and $Y^m_l(\theta,\phi)$ denote the Laguerre polynomials and the spherical harmonics, respectively. The ground state radial wavefunction~\cite{david}
\begin {eqnarray}
\psi_{100} = \frac{1}{\sqrt{\pi a_{\rm B}^3}}\exp\bigg[-\frac{r}{a_{\rm B}}\bigg], \label{eq:18}
\end {eqnarray}
where $L^{2l+1}_{n-l-1}(2r/na_{\rm B}) = L^1_0(2r/na_{\rm B}) = 1$ and $Y^m_l(\theta,\phi) = Y^0_0(\theta,\phi) = 1/\sqrt{4\pi}$. The momentum of the bounded electron in the eigenstate $|\psi_{100}\rangle$ can be obtained from
\begin {eqnarray}
p = \langle\psi_{100}|\textbf{p}|\psi_{100}\rangle = i\hbar\frac{1}{a_{\rm B}}\langle\psi_{100}|\psi_{100}\rangle = i\hbar\frac{1}{a_{\rm B}}, \label{eq:19}
\end {eqnarray}
where $\langle\psi_{100}|\psi_{100}\rangle = 1$. The position of the same electron (with $p = i\hbar/a_{\rm B}$) is given by
\begin {eqnarray}
r = \langle\psi_{100}|\textbf{r}|\psi_{100}\rangle = \int_{0}^{\infty}\bigg[\frac{1}{\sqrt{\pi a_{\rm B}^3}}\exp\bigg(-\frac{r}{a_{\rm B}}\bigg)\bigg]^2r{\rm d}^3r = \frac{3}{2}a_{\rm B}, \label{eq:20}
\end {eqnarray}
where $a_{\rm B} = 4\pi\epsilon_0\hbar^2/m_ee^2$, $\epsilon_0$ is the permittivity of free space, $m_e$ is the mass of electron and $e$ denotes the charge of electron. As anticipated, if the wavefunction is complete (see Eq.~(\ref{eq:18})), the physical quantities ($p$ and $r$) associated to two non-commuting operators ($\textbf{p}$ and $\textbf{r}$) cannot have simultaneous reality. 

For example, $p = i\hbar\frac{1}{a_{\rm B}}$ is not observable simultaneously with $r = \frac{3}{2}a_{\rm B}$ because the momentum is not a real number. This can also be understood physically (see Section 1.6 on page 18 in Ref.~\cite{david}) by noting that the radial wavefunction (Eq.~(\ref{eq:18})) does not represent a proper wave, and therefore, one cannot assign a proper wavelength, $\lambda = h/p$ (de Broglie formula) to Eq.~(\ref{eq:18}), and consequently, the momentum cannot be defined as it should be, and as predicted above. However, the one-peak wave captured by Eq.~(\ref{eq:18}), or Eq.~(\ref{eq:17}) can have a relatively well-defined (relative to $p$) position $r$ for each combination of these quantum numbers ($n$, $l$ and $m$). In other words, since $\textbf{r}$ and $\textbf{p}$ do not commute, the complete wavefunction obtained by solving the Schr$\ddot{\rm o}$dinger equation cannot be used to obtain real values for both $p$ and $r$, which means both $p$ and $r$ cannot be observed simultaneously. Is it possible to guess a `super-complete' wavefunction that can be used to calculate real values for both $p$ and $r$? This is not allowed both mathematically and physically due to $i = \sqrt{-1}$ (see Eq.~(\ref{eq:3})) and $\lambda = h/p$. For example, $r$ requires a one-peak wavefunction while $p$ requires a wavefunction with well-defined wavelength~\cite{david}. However, we can do so (calculate real values for both $p$ and $r$) by writing down (guessing) the incomplete wavefunctions, separately for $r$ and $p$, as originally carried out by EPR~\cite{epr}. But this does not imply (in any way) that we can observe both $p$ and $r$ simultaneously.     

In summary, we have identified the implicit assumption made by EPR to establish or to assert that a complete wavefunction can lead two physical quantities (associated to two non-commuting operators) to have simultaneous reality. The above assertion is based on their hidden \textit{ad hoc} assumption that reads---any wavefunction can be considered as complete if it can be used to solve the Schr$\ddot{\rm o}$dinger equation such that different physical quantities of a given particle or a system can be represented with as many different wavefunctions. For instance, see the discussion after Eq.~(\ref{eq:16}). Their hidden assumption has been exposed to exist by invoking the exactly-solved complete wavefunction for an atomic hydrogen, namely, by defining that a complete wavefunction should remain the same for all measurable physical quantities, otherwise, the wavefunction is incomplete. In other words, what we have shown here are that (i) two physical quantities, which correspond to two non-commuting operators can have simultaneous reality if the wavefunctions are incomplete, and (ii) a complete wavefunction cannot give rise to simultaneous reality for the physical quantities that belong to non-commuting operators. The statements given in (i) and (ii) simply mean that the original EPR paradox (the double negation (or the double-false) explained in the introduction and after Eq.~(\ref{eq:3})) does not exist. Consequently, what we have shown here is that the EPR hidden assumption is logically false with respect to the definition of a complete wavefunction in a real physical system (atomic hydrogen).  

\subsubsection{Entanglement: for particles inside and outside a quantum system}      

The notion of entanglement has been invoked by EPR when they wrote these two equations, Eqs.~(\ref{eq:5}) and~(\ref{eq:7}) because they imply system I, after the interaction is somewhat associated to system II, which was first pointed out by Bohm~\cite{bohm}. To study this effect, Bohm considered a neutral pi meson (pion), which was at rest that decayed into a photon, a positron and an electron, $\pi^0 \rightarrow \gamma + e^+ + e^-$ such that $e^+$ and $e^-$ fly off in the opposite directions. Since the pion has spin zero, one then invokes the conservation of angular momentum to enforce the electron and positron to be in the singlet configuration,
\begin {eqnarray}
\frac{1}{\sqrt{2}}\big(\uparrow_-\downarrow_+ - \downarrow_-\uparrow_+\big), \label{eq:21}
\end {eqnarray}
where $\uparrow_-$, $\downarrow_+$ denote a spin-up electron and a spin-down positron, respectively. The electron and positron that originate from the decayed pion must have opposite spins. The Copenhagen interpretation of quantum mechanics cannot tell us which combination will be measured, but the measurements will be correlated such that on average, one gets $\uparrow_-\downarrow_+$ or $\downarrow_-\uparrow_+$ half the time~\cite{david}. However, the pion itself (before it decays) does not consist of these particles ($e^-$ and $e^+$). It is due to some electromagnetic processes that one obtains a photon, an electron and a positron. The neutral pion actually made up of the combinations of an up-quark ($\texttt{u}$) with an anti up-quark ($\bar{\texttt{u}}$) or a down-quark ($\texttt{d}$) with an anti down-quark ($\bar{\texttt{d}}$)~\cite{djg2}.    

Therefore, the measurements that obtain $\uparrow_-\downarrow_+$ or $\downarrow_-\uparrow_+$ half the time do not prove or disprove the existence of entanglement. For example, one can logically invoke two perfectly valid assumptions, in which, the first one can be used to claim the existence of entanglement, while the other (the second assumption) to counter the first claim. In particular, the first assumption ($\wp$1) reads---both $e^-$ and $e^+$ particles are separated such that their spin configuration is undefined (given by Eq.~(\ref{eq:21})), until one measures the spin. The second assumption ($\wp$2) reads---both $e^-$ and $e^+$ particles are separated with either $\uparrow_-\downarrow_+$ or $\downarrow_-\uparrow_+$ combination. Indeed, $\wp$2 counters $\wp$1, but the Copenhagen interpretation endorses $\wp$1 due to Bohr~\cite{bohr}, although one cannot prove or even show which one of these assumptions is true. The `truth' can only be obtained from `proper' measurements. 

However, $\wp$2 is not related to the hidden variable arguments of Bohm or others~\cite{wer} because we did not impose any additional conditions on $\wp$2, other than what is explicitly stated above. Regardless whether $\wp$2 is true or not, one can go on and develop a theory based on the non-local hidden variable arguments or any other arguments. This is not our objective here. Anyway, the existence of $\wp$1 and $\wp$2, in which one counters the other, and both are valid logically is similar to Russell's paradox in set theory~\cite{monk}. 

To understand why a proper measurement needs to be set up, we need to recall an atomic hydrogen in the ground state. We use atomic hydrogen because the ground state electron and proton are guaranteed to be entangled within the atom and their spin configuration follows Eq.~(\ref{eq:21}) where $\uparrow_+$ and $\downarrow_+$ (now) refer to proton $\tau^+$ (not positron). For example, each time the electron (in the ground state) has spin-up, the proton has spin-down and vice versa. This spin-correlation can be thought of as a time-dependent spin-switch interaction in the electron-proton pair (ground state hydrogen atom), which occurs at timescales known as internal timescales associated to wavefunctions~\cite{ptp}. If this internal timescales are extremely rapid, then it is meaningless to assign spin-up or spin-down to the electron or proton because neither the electron nor proton has a well-defined spin. \textbf{If the spin-correlation (or the spin-switch interaction) timescale is extremely slow (of the order of several days, say), then we can readily assign spin-up (electron) and spin-down (proton) or spin-down (electron) and spin-up (proton). However, these assignments do not imply that we can determine the spin-configuration prior to measurement. Therefore, the quantum theory is self-consistent within a quantum system following Eq.~(\ref{eq:21}).}

However, if the electron is removed from the atom such that the proton is no longer interacting with this unbounded electron, then Eq.~(\ref{eq:17}) does not represent this unbounded electron because $\psi_{nlm} \rightarrow 0$ when $n \rightarrow \infty$. In this case (for a spatially separated electron and a proton), one needs to decide whether $\wp$1 or $\wp$2 is true. Therefore, we need proper experiments to prove or to disprove the existence of entanglement for spatially separated particles. Warning: $n \rightarrow \infty$ does not (in any way) implies the distance between the electron and proton has to be in light years. The distance can be one or two meters. In this case, one can again invoke $\wp$1 (after replacing $e^+$ with $\tau^+$) to claim the existence of entanglement between $e^-$ and $\tau^+$, or conversely make use of $\wp$2 to counter the entanglement. This is why we need a proper measurement to prove $e^-$ and $e^+$ or $e^-$ and $\tau^+$ (they are separated with no interaction between them) is entangled. We cannot enforce $\wp$1 is true just because the Copenhagen interpretation endorses it.

Measuring the spin combination for $e^-$ and $\tau^+$ from an atomic hydrogen is not suitable because one needs to introduce large external disturbances leading to electron excitations before a proper separation can take place. In any case, the atomic hydrogen has served its purpose to show us that within a quantum system (within a bounded $e^-$ and $\tau^+$ system), the quantum theory is self-consistent. We now revert to pion to show that $e^-$ and $e^+$ can be considered entangled even if they are separated a few meters. There should be a device that can switch all spin-up positrons to spin-down positrons. After detecting the spin of the positron, then one detects the spin of the electron. If the electron spin is up for each pair, then $e^-$ and $e^+$ are entangled. This means that we should get spin-up for the electron ($\uparrow_-$) and spin-down ($\downarrow_+$) for the positron all the time if these particles are to be considered entangled. On the other hand, $\wp$2 simply requires the spins for the electrons to be $\uparrow_-$ or $\downarrow_-$ half the time, while for the positrons, one should get $\downarrow_+$ all the time.

In view of Ref.~\cite{gw}, some of you have been carried away into thinking that the only alternative to $\wp$1 is the semiclassical interpretation that is related to insufficient number of entangled pairs being measured (statistically insignificant), without logically acknowledging the existence of a proper alternative to $\wp$1, which is $\wp$2 as explained earlier. In addition, one should be aware here that both $\wp$1 and $\wp$2 are also incompatible with Bell's inequality~\cite{bell}, in which, the local hidden variable arguments cannot be used to discriminate $\wp$1 from $\wp$2. \textbf{The experimental violation of Bell inequalities over long distances reported in Ref.~\cite{gw} does not discriminate between these two valid assumptions, $\wp$1 and $\wp$2. Choosing $\wp$1 without any theoretical or experimental proof to properly rule out $\wp$2 is scientifically not acceptable. In fact, we have proposed additional experimental procedure required to properly measure the spin-configuration such that one can readily rule out $\wp$1 or $\wp$2 or both.} Here, $\wp$2 is also of a quantum mechanical origin. For example, the spins of two particles are entangled within a quantum system, however, these particles are no longer entangled (because they fly off with well-defined spins) when they are spatially separated. Of course, we do not know what is the physical mechanism responsible for the spins to be untangled when they are spatially separated. But we also do not know the physical mechanism responsible for the long-distant instantaneous wavefunction collapse. \textbf{In any case, the wavefunction and spin-eigenstate transformation are physically and logically sound compared to the mysterious wavefunction collapse phenomenon. For example, it makes more sense to think that when a detector detects the electron and its spin, the electron's wavefunction and its spin-eigenstate get transformed due to the interaction with the detector because there is no such thing as the electron or its spin get destroyed after the detection (due to wavefunction collapse). On the contrary, the wavefunction-collapse phenomenon is never proven, nor observed, and the phenomenon is never properly justified to make any sense physically or even logically. But strangely, it was endorsed in Copenhagen under some mysterious arguments due to Bohr and it is still enforced to be the truth.} 

We anticipate that the untangling mechanism could be due to the spin eigenstate transformation, somewhat similar to the wavefunction transformation proven to exist in Ref.~\cite{ptp}. This (untangling mechanism) can be logically understood by noting that if we separate the electron from an atomic hydrogen, then there is this, the wavefunction transformation that reads, $\psi_{nlm} \longrightarrow \psi_{\rm free}$ where $\psi_{\rm free}$ is the wavefunction of an unbounded electron. As a consequence, we can also expect a transformation for the spin eigenstate that reads, $(1/\sqrt{2})(\uparrow_-\downarrow_+ - \downarrow_-\uparrow_+)^{\rm entangled}$ $\longrightarrow$ $(\uparrow_-\downarrow_+)^{\rm untangled}$ or $(1/\sqrt{2})(\uparrow_-\downarrow_+ - \downarrow_-\uparrow_+)^{\rm entangled} \longrightarrow (\downarrow_-\uparrow_+)^{\rm untangled}$. This means that, if the untangling mechanism can be confirmed experimentally between two spatially separated particles (they were entangled prior to separation), then local realism is indeed possible. It is worth noting here that some generalized Bell inequalities have been derived in Ref.~\cite{esp}, which are in agreement with the experimental results, and also support the local realism. Local realism here simply means that the above mentioned spins, when isolated in such a way that they do not interact in any way, then the spin configuration of one particle is independent of the other particle.        

\textbf{Even though we did not provide a proper alternative model theory for the spin eigenfunction transformation to replace the notion of wavefunction collapse phenomenon, but we did provide physically valid arguments on the basis of wavefunction transformation (proven earlier~\cite{ptp}). Moreover, the absence of a proper alternative model theory stated above cannot be used to counter our proven claim `the original EPR paradox does not exist due to a false assumption'. We did provide an alternative model theory that properly expose the process that can kill entanglement when the particles are separated spatially~\cite{ptp}. However, we have made an unproven (but physically valid) proposal where the physical mechanism responsible for spin-untangling is due to spin eigenstate transformation. The generalized proof for particle wavefunction transformation is available in Ref.~\cite{ptp}. The above transformation comes to play when the strength of interaction Hamiltonian changes from one (interacting; both particles exist within a single quantum system) to zero (noninteracting; two particles in two separated and different quantum systems). The above interaction means that the wavefunction does evolve under the spin-interaction Hamiltonian, and if the particles are spatially separated such that they no longer interact, then the spin-interaction Hamiltonian itself ceases to exist. In particular, a quantum state can only remain correlated (entangled) if we assume the spin-interaction (between quantum particles) does not change when the particles are spatially separated. 	According to quantum theory, a quantum state cannot remain the same when the interaction strength changes, the interaction strength cannot remain the same when the particles are spatially separated~\cite{ptp}.} 

\subsubsection{Conclusions}

We have shown that the quantum theory is self-consistent within a quantum system, in accordance with the Copenhagen interpretation of quantum mechanics. In particular, (i) if two operators, $\textbf{A}$ and $\textbf{B}$ commute, then regardless whether the wavefunctions are complete or not, the two corresponding eigenvalues can have simultaneous reality, in addition, (ii) if two operators ($\textbf{A}$ and $\textbf{B}$) are non-commuting, and the wavefunctions are incomplete, then the eigenvalues (corresponding to $\textbf{A}$ and $\textbf{B}$) can have simultaneous reality, on the contrary, (iii) if the wavefunctions are complete, and the two operators are non-commuting, then the two eigenvalues cannot have simultaneous reality, as correctly predicted by the quantum theory. These three points (i-iii) show that the original Einstein-Podolski-Rosen paradox (the double negation given in the introduction) does not exist.  

However, the entanglement controversy raised by Bohm survives for the quantum particles that are non-interacting and are spatially separated. We have interpreted that within a quantum system, the entanglement exist due to some time-dependent spin-switch interaction between these particles and also because of the notion known as internal timescales of the wavefunctions. However, when the entangled particles are separated such that they are not interacting, then the Copenhagen interpretation also claims---the particles are still entangled due to some long-distant instantaneous wavefunction collapse. We found that this claim can only be proven with further measurements.     

If the entanglement phenomenon for the non-interacting particles outside a quantum system is found to be true experimentally, then the quantum theory can be regarded as a supremely exceptional theory. However, if one finds it to be false, then efforts should be made to pave the way for the second interpretation of quantum mechanics on the issue of entanglement for particles spatially separated such that these particles do not interact. Even though these particles were entangled within a quantum system before they split and move away from each other.
  
\section{Acknowledgments}

I am grateful to Madam Sebastiammal Savarimuthu, Mr Arulsamy Innasimuthu, Madam Amelia Das Anthony, Mr Malcolm Anandraj and Mr Kingston Kisshenraj for their financial support and kind hospitality between August 2011 and August 2013. I also would like to thank Mr Mir Massoud Aghili Yajadda (CSIRO, Lindfield) and Mr Alexander Jeffrey Hinde (The University of Sydney) for constantly providing the references for my work. Special thanks to the referee for pointing out the misleading statements.

\end{document}